# Dual Measurements of Temporal and Spatial Coherence of Light in a Single Experimental Setup using a Modified Michelson Interferometer

Mohit Kumar Singh[1,a)] and Shouvik Datta[1,b)]

[1]*Department of Physics & Centre for Energy Science, Indian Institute of Science Education and Research, Pune 411008, Maharashtra, India*

Author(s) to whom correspondence should be addressed:

a) mohitkumar.singh@students.iiserpune.ac.in

b) shouvik@iiserpune.ac.in

**Abstract**

An experimental technique is developed to simultaneously measure both temporal and spatial coherences of a light source by altering a standard Michelson interferometer, which has been primarily used for measuring temporal coherence only. Instead of using simple plane mirrors, two retroreflectors and their longitudinal and lateral movements are utilized to incorporate spatial coherence measurement using this modified Michelson interferometer. In general, one uses Young' double slit interferometer to measure spatial coherence. However, this modified interferometer can be used as an optical setup kept at room temperature outside a cryostat to measure spatio-temporal coherence of a light source placed at cryogenic temperatures. This avoids the added complexities of modulation of interference fringe patterns due to single slit diffraction as well. The process of mixing of spatial and temporal parts of coherences is intrinsic to existing methods for dual measurements. We addressed these issues of spatio-temporal mixing and we introduced a method of 'temporal filtering' in spatial coherence measurements. We also developed a 'curve overlap' method which is used to extend the range of the experimental setup during temporal coherence measurements without compromising the precision. Together, these methods provide major advantages over plane mirror based standard interferometric systems for dual measurements in avoiding systematic errors which lead to inaccuracies especially for light sources with low coherences.

# I. Introduction

Optical interferometry has always been prominently used in important experimental investigations. The famous Michelson interferometer for the search of ether, recent gravitational wave detections in LIGO and Young's double slit experiments for validation of wave-particle duality for photons as well as electrons are some of the important examples. Similar applications of optical interferometry are found to be useful in field of biology[1], astronomy[2], quantum optics[3], holography[4], optical communication[5], surface topography[6] etc. Recently, many-body cooperative phenomena in condensed matter physics of light-matter interactions like excitonic lasing[7–10], polariton lasing[11–17] and Bose-Einstein condensation (BEC) of polaritons[18,19], excitons[20–23] and even photon BEC[24,25] requires the use of optical interferometry to study coherence of emitted light for rigorous experimental verifications. The light emitted by an extended light source preserves the information of phase coherence over certain time delays and over some area of the source, which are known as temporal and spatial coherence of light, respectively. Measurements of the modulus of first order correlation[26] $|g^{(1)}(\Delta r, \Delta \tau)|$ as a function of spatial separation ($\Delta r$) and time delay ($\Delta \tau$) can be used to estimate the spatio-temporal coherence of an extended light source. In fact, simultaneous measurements of spatial and temporal coherences can provide new physical insights of emerging many-body condensed matter physics of these light emitting systems[27,28]. Recently, we also investigated the relation between temporal coherence of emission and auger recombination in III-V semiconductor heterostructure light emitting diode[29]. Most of these many-body cooperative phenomena also occurs at cryogenic temperatures, which requires a cryostat.

Optical power spectrum[30] is connected with the first order correlation function in time through Fourier transform. This can be directly derived from Wiener-Khintchine theorem[31]. Temporal coherence is commonly measured with Michelson and Mach-Zehnder interferometers. However, there are certain conditions which should be considered while measuring temporal coherence such as the longitudinal spatial coherence[32] which can limit the temporal coherence during its measurement. These conditions are discussed in details by Ryabukho et. al[32–36] and Abdulhalim[37,38]. Since we are dealing with interference of light spots sufficiently far from the source, we do not see any significant involvement of longitudinal spatial coherence during the measurement of temporal coherence. For spatial coherence measurement, Young's double slit interferometer[39] is typically used. Usually, double slit method yields very low intensity output due to use of narrow slits and it is difficult to use this method when the intensity of light emission is already low. Individual slits also produce their own diffraction patterns which should be filtered/subtracted from double slit interference pattern to correctly measure $|g^{(1)}(\Delta r)|$. Additionally, it is not easy to place double slits with variable slit widths and slit separations within a cryostat in close proximity of the sample for spatial coherence measurements at cryogenic temperatures.

Moreover, using two separate experimental setups for any generalized measurements of first order spatio-temporal correlation $|g^{(1)}(\Delta r,\Delta \tau)|$ can be inconvenient from the point of view of experimentation. Hence, modified Young's interferometer[34,40,41] and several combinations of Michelson[27,42–44] and Mach-Zehnder interferometers[45,33] were used to perform combined measurement of spatial and temporal coherence. A single setup is more convenient to measure both types of coherence but there are instrumentational challenges as interferometry is very sensitive on experimental conditions. One such major challenge is to measure spatial and temporal coherence separately using a single interferometer. The effect of longitudinal spatial coherence in temporal coherence measurement due to properties of the light source has been studied[32,34,36–38], but mixing of these coherences due to instrumentational designs is hardly studied. For example, an unwanted spatio-temporal phase always adds up due to tilting of interfering beams. This type of spurious angular change(s) may occur due to the design and technique of a particular instrument like changing angle of beam for partial overlap, use of one lens for two spatially shifted beams, optical misalignments etc. Some of these temporal phases can add systematic errors during measurement of spatial coherence. In some dual measurement setups reported earlier[42,45,47], these unwanted temporal phases are not considered which might lead to additional errors. However, with design modication[33], some of these systematic errors can be reduced but an additional method is required, which assumes that correlation function is not wholly dependent upon the design to minimize the mixing of these phases. Here, we use a technique of 'temporal filtering' method to maximally avoid such unwanted mixing during the measurement of spatial coherence.

It is often required to notice changes in spatio-temporal coherence due to the onset of some physical phenomenon like lasing[8,14–16], BEC[19,48] as mentioned above. There, we require an instrument whose range can be extended to measure these substantial changes. So, during the measurement of temporal coherence, we use a 'curve overlap' method to extend the range of the setup while maintaining its precision. Moreover, this modified Michelson interferometer can offer the technical convenience for such kind of instrumentations where one may not be able to place adjustable double slits for measuring spatial coherence (say) within a cryostat.

We divide this report in four sections. In the section II, we start with a brief introduction of $g^{(1)}(\Delta r, \Delta \tau)$ and its measurement methods. In section III, we first explain the complete experimental details of the instrumentations and then further divide this section into two subsections. Standard operating procedure of Michelson interferometer, data acquisition and analyses methods are described in the subsection on 'Temporal correlation measurement'. The technique of 'curve overlap' and results are also discussed in this subsection. In the second subsection on 'Spatial correlation measurement using the modified Michelson interferometer', we explain the modifications in acquisition and analyses of Michelson interferometer to measure spatial coherence as well. At the end of this subsection, we discuss the results of this modified setup. Finally, in section IV, we summarize our conclusions.

## II. First order correlation function $g^{(1)}(\Delta r, \Delta \tau)$ from optical interference pattern

The first order correlation function between two optical fields is defined as[49]

$$g^{(1)}(r_1,r_2,t_1,t_2) \equiv g^{(1)}(\Delta r, \Delta \tau) = \frac{\langle E(r_1,t_1)E^*(r_2,t_2)\rangle}{\sqrt{\langle |E(r_1,t_1)|^2\rangle \langle |E(r_2,t_2)|^2\rangle}} \quad (1)$$

where $E(r_1,t_1)$ and $E(r_2,t_2)$ are two interfering electric fields of light at position $r_1$ and $r_2$ at time $t_1$ and $t_2$, respectively. It is also defined in relative terms as $\Delta r = r_1 - r_2$ and $\Delta \tau = t_1 - t_2$. The $\Delta \tau$ is the temporal delay between two interfering fields at one spatial position. In Eq. (1), when space coordinates are kept constant as $r_1 = r_2$, then correlation measured between $t_1$ and $t_2$ is called temporal correlation function or $g^{(1)}(\Delta \tau)$. Similarly, when time coordinates are kept constant as $t_1 = t_2$, $g^{(1)}(\Delta r)$ is called spatial correlation function at one point of time. Here $\Delta r$ is spatial separation between two interfering fields.

The information of spatio-temporal coherence is estimated by measuring the modulus of $g^{(1)}(\Delta r, \Delta \tau)$ function. The $|g^{(1)}(\Delta r, \Delta \tau)|$ is strictly positive quantity ($0 \leq |g^{(1)}(\Delta r, \Delta \tau)| \leq 1$). While using any optical interferometer, the method to measure $|g^{(1)}(\Delta r, \Delta \tau)|$ is to interfere two light fields and analyze their interference pattern. Interference of these light fields forms two dimensional (2D) interference pattern on a screen and one single line/axis (1D) from the pattern is chosen for analysis. This 1D interference pattern can be extracted by modulating[45,47] the spatio-temporal delays between two optical fields or it can also be directly extracted from the 2D fringe pattern[50]. In either case, a small phase difference between two fields is introduced to produce the fringe pattern. It is kept much smaller than the operational phase delays introduced while moving the retroreflectors during the actual measurements. This also provides an instrumentational lower bound on $|g^{(1)}(\Delta r,\Delta\tau)|$, which can be measured without mixing temporal and spatial parts. Anyway, all these time delays (or path differences) between two interfering fields must also remain smaller than the coherence time (or lengths) of sources. In this work, we use 2D interference pattern and select one axis to extract fringe pattern for analysis.

The fringe intensity is either fitted with interference formula given below or the simple visibility[31,49] of fringes is measured to get the value of $|g^{(1)}(\Delta r,\Delta\tau)|$. The 1D interference formula along one axis can be expressed as[51]

$$I_{int}(x) = I_1(x) + I_2(x) + 2|g^{(1)}(\Delta r,\Delta\tau)|\sqrt{I_1(x)I_2(x)}\cos(k_x x - \varphi) \qquad (2)$$

where x is the 1D coordinate along which the variation of interference pattern is measured on CCD/screen. $I_{int}(x)$ is intensity due to interference of two fields $E(r_1,t_1)$ and $E(r_2,t_2)$, $I_1(x)$ and $I_2(x)$ are intensities of individual field along the same axis. $k_x x = (\vec{k_1} - \vec{k_2}) \cdot \vec{x}$, where $\vec{k_1}$ and $\vec{k_2}$ are wave vectors of $E(r_1,t_1)$ and $E(r_2,t_2)$, respectively, $(\vec{k_1} - \vec{k_2}) \cdot \hat{x} = k_x$ and $\varphi$ is a constant phase factor. Once $I_{int}(x)$, $I_1(x)$ and $I_2(x)$ are known, the $|g^{(1)}(\Delta r,\Delta\tau)|$ can be estimated from the above relation. Here, $\Delta r$ and $\Delta\tau$ are experimental variables which are set by instrument. The precision of these input variables is instrument dependent.

### III. Modified experimental setup to determine temporal and spatial correlation

Schematic diagram of the complete experimental setup is given in Fig. 1(a). We are using a standard Michelson interferometer setup. However, the usual plane mirrors are replaced with two silver coated retroreflector[52,53] mirrors $M_1$ and $M_2$ procured from Holmarc, India and one CCM1-BS013/M non-polarizing 50:50 beam splitter (BS) from Thorlabs is used. The distances of $M_1$ and $M_2$ from center of beam splitter are $L_1$ and $L_2$, respectively. The retroreflectors are also effective in blocking any back-reflection into the light source, which can otherwise create instabilities and affect the coherence of the light source. Double retroreflector design thus prevents any unwanted optical effects such as image flipping (if only one retroreflector is used) and mode hopping, intensity fluctuations, coherence oscillations[54,55] etc. due to absence of any unwarranted positive feedback from the source during the measurement.

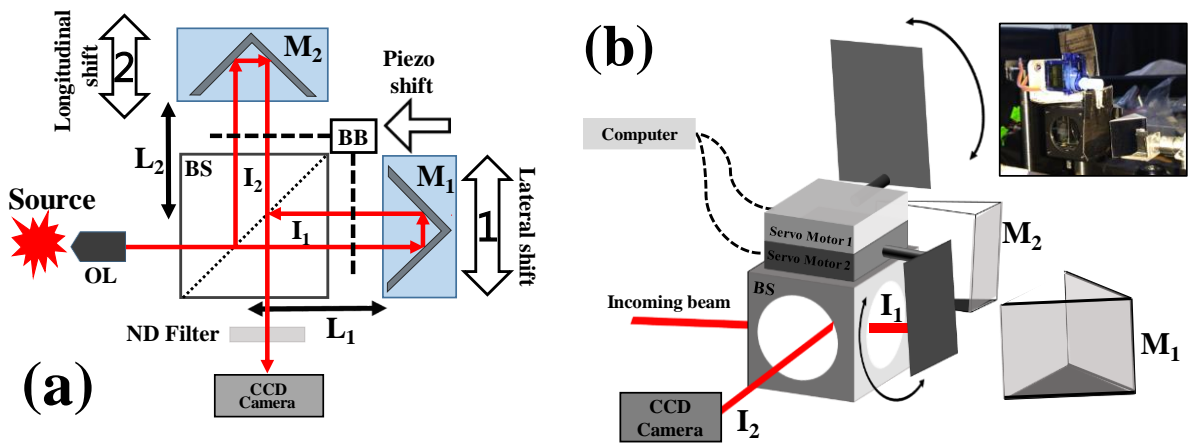

**Fig. 1.** (a) Schematic diagram of experimental setup based on Michelson interferometer. $M_1$ and $M_2$ are two retroreflectors and BS is 50:50 beam splitter. $M_1$ and $M_2$ are installed on linear stages to move in the direction shown by arrow 1 and 2, respectively. $M_1$ is attached to a Piezo actuator which moves in the direction labelled as 'Piezo shift'. Source is collimated by objective lens (OL) of 10X magnification. The light beam split as $I_1$ and $I_2$ are reflected back to generate an interference pattern on the plane of the CCD. The distances of $M_1$ and $M_2$ from centre of BS are $L_1$ and $L_2$, respectively. (b) Schematic diagram of beam blocker over BS. In the inset, a real image of BB is shown.

The light source is an AlGaInP based multi-quantum well commercial laser diode HL6358MG (Thorlabs) which emits at a wavelength of 639 nm. The bias to laser diode is given by ITC4001 Controller (Thorlabs). The laser diode has lasing threshold at 30 mA but whole experiment is done well below lasing threshold, at 1 mA. The reason for using laser diode in a non-lasing mode is because we are interested in making an experimental setup which can be used for measuring small and variable coherence of a light source. The source is collimated by using a 10X objective lens. $M_1$ is attached to a piezo actuator PAS005 (Thorlabs). Piezo can move up to ~ 16 µm. The direction of piezo displacement is shown by arrow labelled as 'Piezo shift' in Fig. 1(a). The Piezo and $M_1$ combination is fitted on a kinematic alignment mount for beam alignment. This kinematic mount is installed on a linear stage which moves in the direction indicated by arrow 1 (lateral shift). This stage has precision of 10 µm and it can move up to 25 mm. Similarly, the other retroreflector $M_2$ is also fitted with a similar kinematic mount (without Piezo actuator) on a linear stage. This stage of $M_2$ moves in direction shown by arrow 2. We call it as longitudinal shift. These manual stages are moved by stepper motors and are remotely controlled using a computer.

Beams reflected from $M_1$ and $M_2$, indicated as beam $I_1$ and $I_2$, are made to interfere at the CCD camera of BC106N-VIS/M beam profiler (Thorlabs). The sensor of CCD has 1360 × 1024 pixels and sensor area is ~8.77 × 6.60 mm$^2$. The distance on CCD sensor are recorded in terms of pixel number. These pixel numbers are converted to distance (µm or mm) and are defined in X-Y coordinates (CCD sensor coordinates). In front of the CCD camera, we use neutral density (ND) filters from Newport to reduce the light intensity by 0.3 dB to 4 dB. This is to protect camera sensor from any intense light and subsequent overloads. Over the beam splitter, we have installed a custom made beam blocker (BB) which use two servo motors and two opaque sheets, as shown in Fig. 1(b), to block either arm of the interferometer. It blocks $I_1$ beam while letting $I_2$ to pass and vice versa to measure $I_1$ and $I_2$ separately. In the inset of the same figure, the real image of beam blocker is shown. This whole setup (except the light source and CCD camera) is assembled on a single optical breadboard (25cm × 25cm). After initial alignment, we can easily relocate this portable setup to use it in different experiments. Initial alignment of whole setup is done with a He-Ne Gas laser.

**A. Temporal correlation measurement**

The lateral shift of $M_1$ can move $I_1$ beam horizontally on the screen of CCD. For measurement of temporal correlation, in Fig. 1(a), the linear stage of $M_1$ is locked at a position in lateral direction such that $I_1$ and $I_2$ beams are completely overlapping with each other. Although, $M_1$ is locked for any lateral shift, but piezo actuator attached with $M_1$ moves in the direction shown in Fig. 1(a). Hence, piezo movement of $M_1$ is to produce fine temporal delay while longitudinal movement of $M_2$ is to produce coarse temporal delays between $I_1$ and $I_2$ beams. The temporal delay $\Delta\tau = 2*(L_2 - L_1)/c$ where $(L_2 - L_1)$ is the length difference between two arms of the interferometer and c is speed of light. Linear stage of $M_2$ produces $\Delta\tau \sim 66$ fs when moved by 10 μm. Piezo steps are not constant but change quadratically as a function of piezo voltage. It can cover total $\Delta\tau \sim 108$ fs in 40 steps.

*1. Data Acquisition Method*

We collect 2D images of $I_1$, $I_2$ and $I_{int}$ along with camera parameters such as gain, exposure-time, saturation and sensor temperature for each image. The camera is used in auto-exposure mode to automatically maintain the optimum values of gain and exposure time depending upon the light intensity. Moreover, we used LabView based data acquisition algorithm to select images which have approximately similar values of these camera parameters. Detailed flow-chart for data acquisition is shown in Fig. 2. For initial alignment, we adjust $M_1$ and $M_2$ stages and kinematic mounts for total overlap of $I_1$ and $I_2$ beam spots on CCD. Then, we save image of $I_1$ by blocking the $I_2$ beam. We mark the position of $I_1$ spot by using four cursor lines on the CCD output screen. Then, we block $I_1$ while opening $I_2$ and adjust the spot position of $I_2$ beam according to the set cursor lines to ensure the total beam overlap. We save the image of $I_2$ after making sure that both beam spots are properly overlapping. For saving the image of $I_{int}$, both the paths are kept open. Images of $I_{int}$ are saved at each of the 40 steps of the piezo movement during a single run. For the coarse temporal delay, the linear stage of $M_2$ is set to the next step and piezo steps are again repeated. Since total fine delay is almost twice the coarse delay, some portion of data is repeated at each linear stage step.

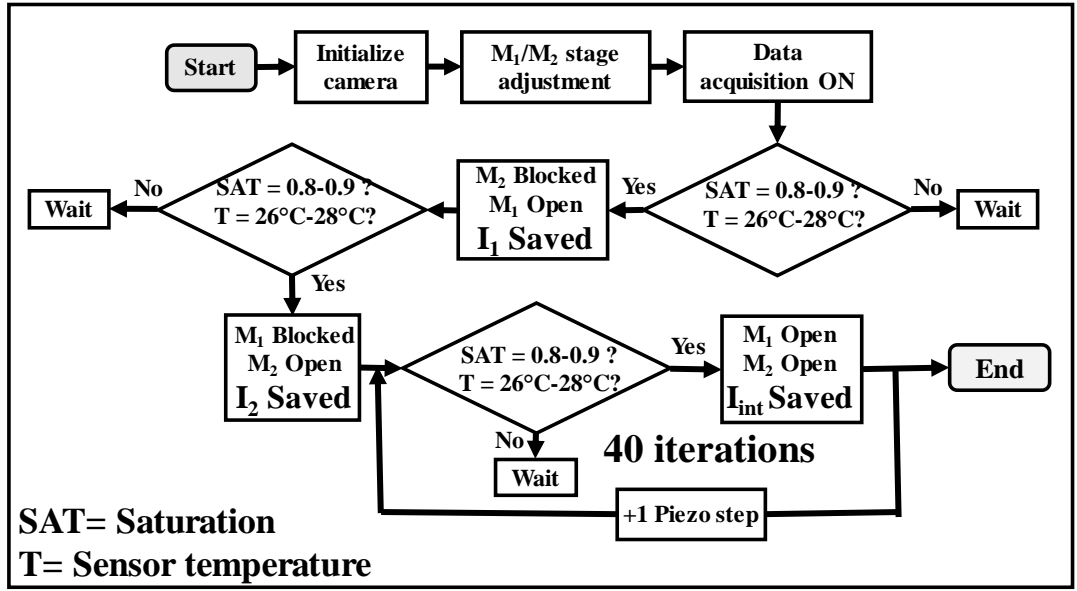

**Fig. 2.** Flow-chart of data acquisition method used in the experiment. Saturation value (SAT), and sensor temperature (T) of camera are kept in narrow range while saving 2D images to minimise the impact of camera parameters. Using beam blocker, $I_1$ and $I_2$ are saved by blocking $M_2$ and $M_1$, respectively. For each step of stage $M_1$ (or $M_2$ for the case of spatial correlation measurement), $I_1$ and $I_2$ are saved and images of $I_{int}$ are also saved at each of the 40 steps of the piezo.

## 2. Data Analysis Method

The collected 2D images are saved as 2D arrays. The scaling is done by dividing these 2D-arrays by their corresponding gain and exposure time values. From the 2D image of $I_{int}$, one axis is chosen to extract the 1D data. One such image is given in Fig. 3(a) and chosen axis is indicated by σ-axis at the centre of $I_{int}$ image shown by black dashed line. X-Y axis are CCD sensor coordinates along the horizontal and vertical directions. The choice of σ-axis as perpendicular to X-axis is mainly to maintain the similarity with measurements of spatial coherence along the X-axis itself. The intensity values over this axis are extracted from images of $I_{int}$, $I_1$ and $I_2$ as $I_{int}(\sigma)$, $I_1(\sigma)$ and $I_2(\sigma)$. We take 10 pixels on either side of σ-axis and average these intensities. This averaging is done to avoid any pixel related errors. Estimated values of $I_{int}(\sigma)$, $I_1(\sigma)$ and $I_2(\sigma)$ are shown separately in Fig. 3(b).

To calculate $|g^{(1)}(\Delta\tau)|$ from these data, we rearrange Eq. (2) and replace the 'x' coordinates with our chosen σ-coordinates to get the cosine function as -

$$\frac{I_{int}(\sigma)-I_1(\sigma)-I_2(\sigma)}{2\sqrt{I_1(\sigma)I_2(\sigma)}} = |g^{(1)}(\Delta\tau)|\cos(k_\sigma \cdot \sigma - \varphi) \equiv S(\sigma) \qquad (3)$$

Here, $S(\sigma)$ is acquired by using intensities of $I_{int}(\sigma)$, $I_1(\sigma)$ and $I_2(\sigma)$ in Eq. (3). If all intensities are corrected for the above mentioned scaling, $S(\sigma)$ should mimic the cosine function and the amplitude of this function should be equal to $|g^{(1)}(\Delta\tau)|$. The $S(\sigma)$, estimated for $I_{int}(\sigma)$, $I_1(\sigma)$ and $I_2(\sigma)$ of Fig. 3(b), is shown in Fig. 3(c). As mentioned in section II, $|g^{(1)}(\Delta\tau)|$ is strictly positive ($0 \leq |g^{(1)}(\Delta\tau)| \leq 1$). Also, dark currents of CCD and other electronics can affect the estimate of $S(\sigma)$. Therefore, instead of fitting $S(\sigma)$ with a simple cosine function, we modify the fitting function as

$$\text{Fitting function of } S(\sigma) = S_0 + \exp(b)\cos\left(\frac{\sigma-\sigma_0}{d}\right) \qquad (4)$$

where $|g^{(1)}(\Delta\tau)| = \exp(b)$ fullfils the condition of $|g^{(1)}(\Delta\tau)| > 0$ and $S_0$ is an offset which shows effect of dark currents or systematic errors in calculation of $|g^{(1)}(\Delta\tau)|$. The $d$ and $\sigma_0$ depend upon values of $k_\sigma$ and $\varphi$. We select the central section from $S(\sigma)$ to include at least 2-3 oscillations for cosine fitting shown as red dashed rectangle in Fig. 3(c).

In Fig. 3(a), from CCD sensor coordinate axis X ~ 3.87 mm to X ~ 6.13 mm (as shown by red dashed lines), different σ-axis with separation of ~0.06 mm (10 pixels) are chosen to estimate $|g^{(1)}(\Delta\tau)|$. Here, we do not take the average over 21 adjacent pixels. These $|g^{(1)}(\Delta\tau)|$

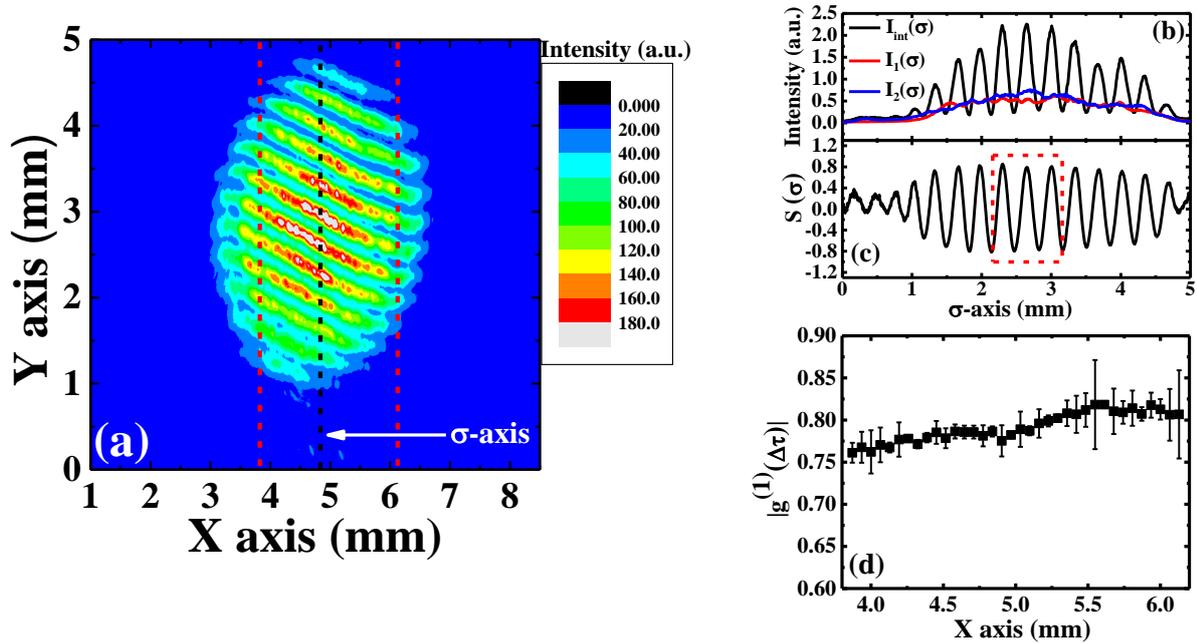

**Fig. 3.** (a) Interference image due to complete overlap of $I_1$ and $I_2$ beams. X-Y are camera sensor coordinates. Temporal correlation measurements are done maintaining this overlap condition. The σ-axis is in the centre of interference pattern indicated by black dashed line. (b) $I_1(\sigma)$, $I_2(\sigma)$ and $I_{int}(\sigma)$ are extracted from interference image along σ-axis. (c) $S(\sigma)$ is calculated using Eq. (3). Different σ-axes are chosen from sensor coordinate X ~ 3.87 mm to X ~ 6.13 mm indicated by red dashed line on (a). The value of $|g^{(1)}(\Delta\tau)|$ is calculated along these σ-axes and are plotted against sensor X coordinates in (d). Error bars are offset $S_0$ from Eq. (4).

are plotted in Fig. 3(d) with offset $S_0$ as error bar for comparisons. It shows that $S_0$ are small compared to $|g^{(1)}(\Delta\tau)|$ values. Most importantly, it is seen that small variations in the selection of σ-axis do not affect the results. The mean value of all the $|g^{(1)}(\Delta\tau)|$ is 0.79 and standard

error is 0.01. Hence, the methods adopted for data acquisition and analyses are robust against such small displacements of σ-axis.

### 3. *Discussion of Experimental results on Temporal Coherence Measurement*

To calculate $|g^{(1)}(\Delta\tau)|$, we make two beams to overlap completely reducing $\Delta r \sim 0$ mm as mentioned in section III.A.1. We set linear stage of $M_2$ at a starting point $\tau_0$ and then another 40 piezo steps are taken in the direction of 'piezo shift' (see Fig. 1(a)). Then, $M_2$ stage is moved by one step (~66 fs) to the point $\tau_0$ + 66 fs and again 40 piezo steps are taken. This way, the piezo and linear stage can provide fine and coarse delay ($\Delta\tau$), respectively. In Fig. 4(a), multiple curves are shown and each individual curve is $|g^{(1)}(\Delta\tau)|$ calculated for piezo steps while different curves are for steps of $M_2$ stage. The temporal position of $M_2$ stage is indicated by [$\tau_0$ + 1 x 66] fs, [$\tau_0$ + 2 x 66] fs etc. As we already mentioned, step of $M_2$ stage is of ~66 fs and 40 steps of piezo covers ~108 fs. Hence we see significant overlapping between adjacent curves of $|g^{(1)}(\Delta\tau)|$. We translate these individual curves to overlap into a single curve as shown in inset of Fig. 4(a). The maximum overlapping condition is decided when Δg is minimum for overlapping portion of two curves. This Δg is defined as,

$$\Delta g = \frac{\sum_i \{(g_n^{(1)})_i(\Delta\tau) - (g_{n+1}^{(1)})_i(\Delta\tau \pm t)\}}{\sum i} \quad (5)$$

where i = number of overlapping point in curve. $(g_{n+1}^{(1)})_i$ are points of (n+1)$^{th}$ curve of $M_2$ stage which overlap with adjacent (n)$^{th}$ curve. The (n+1)$^{th}$ curve is translated in $\Delta\tau$ by the unit of t = N x (minimum step of piezo) where N = 1,2,3… . The overlapped curves are then stitched together to produce single curve of $|g^{(1)}(\Delta\tau)|$ as shown in Fig. 4(b). The main benefit of the curve overlap method is that it extends the range of measurement while retaining the precision. We mentioned few physical phenomena in section I where coherence of the source changes and this curve overlap method can be very helpful to measure those changes in coherence. Moreover, good overlap also indicates good reproducibility and reliability of data. The error bars used in the plot are the 95% confidence interval (CI). The CI is calculated by multiplying standard error to student-t inverse cumulative distribution function. The standard error is calculated from covariance matrix which is generated for the fitting parameters of Eq. (4). The $S_0$ values which represent the offset due to dark currents of CCD are smaller in comparison with the 95% CI. The instrumental errors/uncertainties in $I_{int}(\sigma)$, $I_1(\sigma)$ and $I_2(\sigma)$ due to CCD propagate to the error of estimated $|g^{(1)}(\Delta\tau)|$ (see Eq. (3)). This is also small in comparison with the 95% CI. Therefore, only the CI is used as error bars. Each curve overlaps

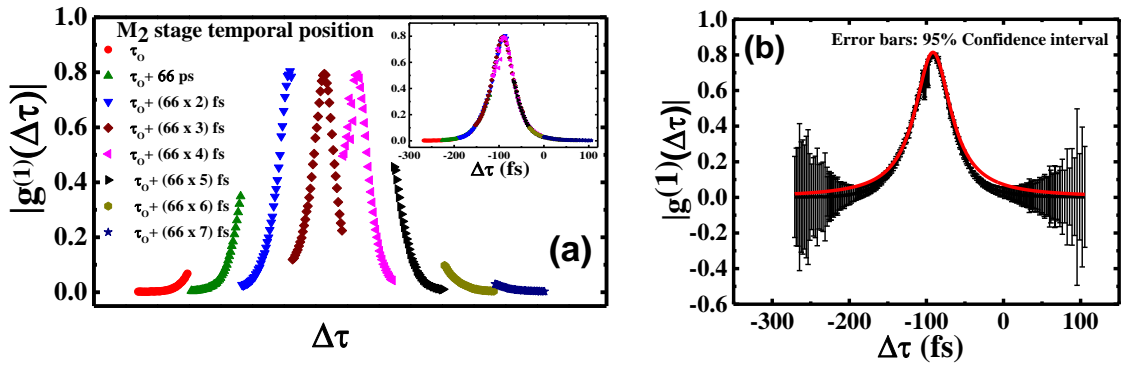

**Fig. 4.** (a) shows the estimated $|g^{(1)}(\Delta\tau)|$ plotted agaisnt fine motion of piezo ($\Delta\tau$) at different $M_2$ stage temporal position. Plots of different $M_2$ stage are manually segregated along $\Delta\tau$ axis for clarity. Therefore, scale markers along $\Delta\tau$ are not given. $M_2$ stage has coarse temporal delay of $\Delta\tau \approx 66$ fs and total piezo temporal delay is ~108 fs. Therefore, there are significant overlapping between adjacent curves. Inset of (a) shows the overlapped curve by translating indivudual curves on $\Delta\tau$ axis. (b) Overlapped curves are then stitched to a single curve and fitted with Lorentzian function. Fitting is shown as red line. Error bars are of 95% confidence interval.

with two immediate adjacent curves on either side by ~60% and two more adjacent curves by ~20%. Any misaligned overlap would have suggested that either the source is not stable or there are changes due to movement of $M_2$ stage. However, we see reasonably good extent of curve overlaps in inset of Fig. 4(a). This indicate that data is reproducible even after movements of linear stage $M_2$ and the beam alignment is maintained to fulfil the condition of total beam overlap. These matching overlaps further validate our instrumentation and the method of analyses as mentioned above. The full curve of $|g^{(1)}(\Delta\tau)|$ is then fitted with Lorentzian function,

$$|g^{(1)}(\Delta\tau)| = g_0 + \frac{2(A)}{\pi} \left( \frac{w}{4(\Delta\tau - \Delta\tau_{center})^2 + w^2} \right) \qquad (6)$$

where $g_0 (\geq 0)$ is offset, w is FWHM, A is area and $\Delta\tau_{center}$ is centre of curve on $\Delta\tau$ axis. $g_0$ is 0.0, $\Delta\tau_{center}$ is -91 fs, A is 73 $fs^2$ and FWHM is 57 fs. $R^2$ value of fitting is 0.979. The approximated coherence time ($\tau_c$) of the source is FWHM the curve of $|g^{(1)}(\Delta\tau)|$. The error in FWHM due to fitting is 1 fs. The highest step size of our piezo is ~3 fs. So this step size is used as scale of error in the measurement of $\Delta\tau$. Therefore, the estimated coherence time ($\tau_c$) is 57 ± 3 fs. We also calculate (95% CI of $|g^{(1)}(\Delta\tau)|$) /$|g^{(1)}(\Delta\tau)|$ to be < 0.1 for values of $|g^{(1)}(\Delta\tau)| > 0.17$. The same value is < 0.5 for $|g^{(1)}(\Delta\tau)| > 0.05$. We also calculated the Fourier transform[30] of the measured optical power spectra of the source and then estimated the coherence time as ~62 fs from its FWHM using Wiener-Khintchine theorem[31]. Therefore, it shows that two independent measurements estimate approximately similar values of $\tau_c$. This establish that our Michelson interferometer is working properly.

**B. Spatial correlation measurement using the modified Michelson interferometer**

In the modified Michelson inteferometer, when $I_1$ and $I_2$ beams partially overlap on CCD plane, the resulting interference contains the information about spatial correlation. To understand this, we draw the schematic of the original light beam's cross sectional image before it passes through beam splitter in Fig. 5 (left). The different dotted and dashed lines indicate different parts of the beam spot and identical lines among these are seperated by distance $\Delta r$. The schematics of the reflected beams from $M_1$ and $M_2$ on CCD are shown in Fig. 5 (right). If the speration between two beams ($I_1$ and $I_2$) is $\Delta r$, the identical lines will be in complete overlap. The interference generated by this partial overlap ($\Delta r$) can be used to measure spatial coherence if $\Delta\tau\sim 0$.

In previous experiments, the partial overlap is achieved by changing the reflection angle of one of the mirrors[45] of interferometer or by spatially inverting one of the beams[27,28,42,44,56] with only one retroreflector. The change in reflection angle and merging of spatially shifted beams also change the angle between overlapping beams. This angular deviation leads to unwanted temporal phase difference. To avoid this, linear deviation of the beam instead of angular deviation is used to achieve partial overlap[47] while measuring spatial coherence. In Fig. 1(a), $I_1$ beam is shifted by laterally shifting the linear stage of $M_1$ in direction as indicated

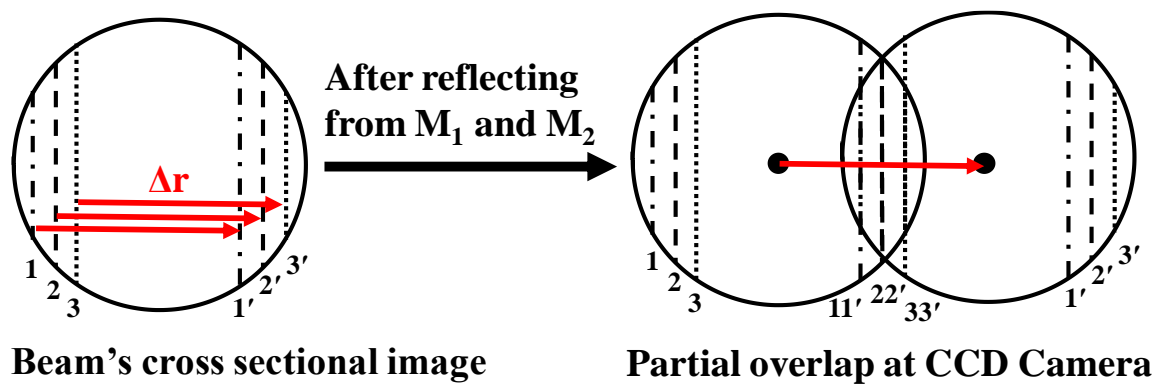

**Beam's cross sectional image**      **Partial overlap at CCD Camera**

**Fig. 5.** (Left) Schematic cross-sectional image of beam originating from objective lens before entering to beam splitter. Identical dotted and dashed lines (1 and 1', 2 and 2', 3 and 3') are $\Delta r$ distance apart in the same beam. (Right) Partially overlapping image of $I_1$ and $I_2$ after reflecting back from $M_1$ and $M_2$. The beams are $\Delta r$ distance apart and identical dotted and dashed lines overlap. The resulting interference pattern due to this partial overlap can be used to measure spatial coherence, if $\Delta\tau\sim 0$. This ideal schematic, however, does not include the small phase difference introduced to obtain the fringe pattern in the first place.

by arrow 1. M₂ is locked in position of Δτ~0. The lateral shift of M₁ shifts the I₁ beam without any angular deviation, hence, it prevents any additional Δτ. We use CCD sensor coordinates to measure the lateral shift of I₁ by noticing the shift in specific marker on beam when M₁ is being shifted. The pixel size of the CCD sensor is 6.45 µm, which limits the spatial resolution.

In Fig. 6(a), 2D image of the intereference pattern $I_{int}$ is shown for partial seperation of $I_1$ and $I_2$ by Δr~2.8 mm on CCD sensor coordinates. The σ-axis is chosen at the middle of fringe pattern and perpendicular to the direction of spatial translation along X, which is parallel to the lateral shift of M₁ along the horizontal plane. The measured $I_{int}(σ)$, $I_1(σ)$, $I_2(σ)$ are plotted in Fig. 6(b). The S(σ) is calculated from these intensities using Eq. (3) and S(σ) shows a clear cosine pattern in Fig. 6(c). The central portion of S(σ) is taken for fitting with Eq. (4) as shown by red dashed rectangle and the amplitude of this cosine function is estimate of $|g^{(1)}(Δr)|$. After the partial overlap, one can also use visibility to measure $|g^{(1)}(Δr)|$ as,

$$\text{Visibility} = \frac{I_{max} - I_{min}}{I_{max} + I_{min}} = 2|g^{(1)}(\Delta r)| \frac{\sqrt{I_1(\sigma) I_2(\sigma)}}{[I_1(\sigma)+I_2(\sigma)]} \quad (7)$$

where $I_{max}$ and $I_{min}$ are successive maximum and minimum intensity of interference pattern ($I_{int}(\sigma)$) on central portion of σ-axis, $I_1(\sigma)$ and $I_2(\sigma)$ are intensity of individual beam for the same portion. This formula, however, requires the strict condition of $I_1(\sigma) = I_2(\sigma)$ for visibility = $|g^{(1)}(\Delta r)|$. However, the light beam may not be symmetric in intensity around the center for many sources, therefore $I_1(\sigma)$ and $I_2(\sigma)$ can have different intensities. Therefore, we avoid using visibility as a measure of $|g^{(1)}(\Delta r)|$ and individually take $I_1(\sigma)$ and $I_2(\sigma)$. We use Eq. (3) and (4) to get $|g^{(1)}(\Delta r)|$. The detailed data acquisition method is already given in section III.A.1 where instead of total overlap, we only partially overlap light beams to introduce $\Delta r$.

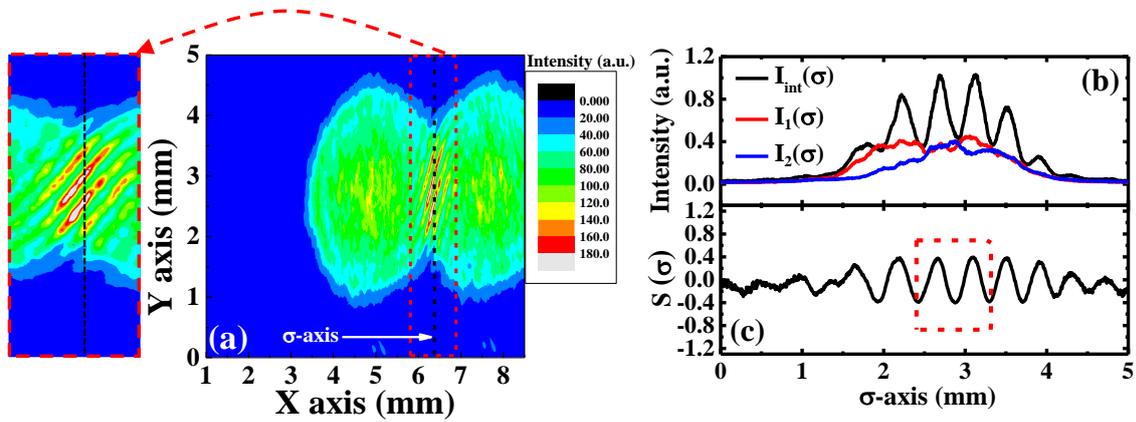

**Fig. 6.** (a) Interfere image due to partial overlap of I₁ and I₂ beams. Beams are separated by ~2.8 mm on CCD sensor coordinates. The σ-axis is in the centre of interference pattern indicated by black dashed line. Magnified image of interference pattern is shown in red dashed lined box (left). (b) $I_1(σ)$, $I_2(σ)$ and $I_{int}(σ)$ are extracted from interference image along σ-axis. (c) $S(σ)$ is estimated using Eq. (3).

### 1. Data Acquisition, Analysis and Results on Spatial Coherence Measurement

As discussed above, to eliminate the involvement of unavoidable temporal phase, we used the lateral shift of $M_1$. However, there can be other source of unwanted temporal phase as well which we discussed in section I. We use a technique called 'temporal filtering' in spatial coherence measurement filter out any temporal phase. Here, we use piezo to make temporal correlation scan for every step of $M_1$. At every beam separation ($\Delta r$), piezo takes 40 temporal steps to measure $|g^{(1)}(\Delta r, \Delta \tau)|$. The $\Delta r$ is introduced by lateral shift of $M_1$ and $\Delta \tau$ is due to piezo shift. These $|g^{(1)}(\Delta r, \Delta \tau)|$ at different $\Delta r$ values are plotted in Fig. 7(a). The maximum value of $|g^{(1)}(\Delta r, \Delta \tau)|$ of every curve indicates the position of where $\Delta \tau = 0$. By picking these points $|g^{(1)}(\Delta r, \Delta \tau=0)|$ the involvement of any temporal phase is eliminated. $|g^{(1)}(\Delta r, \Delta \tau=0)|$ are plotted against $\Delta r$ in Fig. 7(b). This way, we are able to filter contributions of $\Delta \tau$ in spatial coherence measurement and $|g^{(1)}(\Delta r)|$ is estimated. The curve of $|g^{(1)}(\Delta r)|$ is fitted with Lorentzian function as given in Eq. (6) where $\Delta \tau$ and $\Delta \tau_{center}$ is replaced by $\Delta r$ and $\Delta r_{center}$. $R^2$ value of fitting is 0.994. $g_0$ is 0.2, $\Delta r_{center}$ is 0.01 and A is 3.8 mm$^2$. FWHM is 3.8 mm with standard error of fitting ~0.2 mm. Here, the standard error of fitting is greater than the instrumental precision of $\Delta r$, hence it is being used as uncertainty of the estimated FWHM. The FWHM length can be called as the spatial separation on beam spot up to which phase correlation is significant. Since we are using magnification of 10X, on the emitting surface of source, this length is 380 ± 20 µm. This is called 'spatial coherence length' ($\Delta r_c$) on the surface of emission.

The reason why 'spatial filtering' is not used during temporal coherence measurement is because any significant spatial shift of two beam spots can be detected on CCD screen. We explained in section III.A.1 that four cursor lines are used to mark the positon of individual beam spot to avoid any spatial shift. On the other hand, if additional temporal phase is introduced through instrumentations, it cannot be easily detected from the beam overlap.

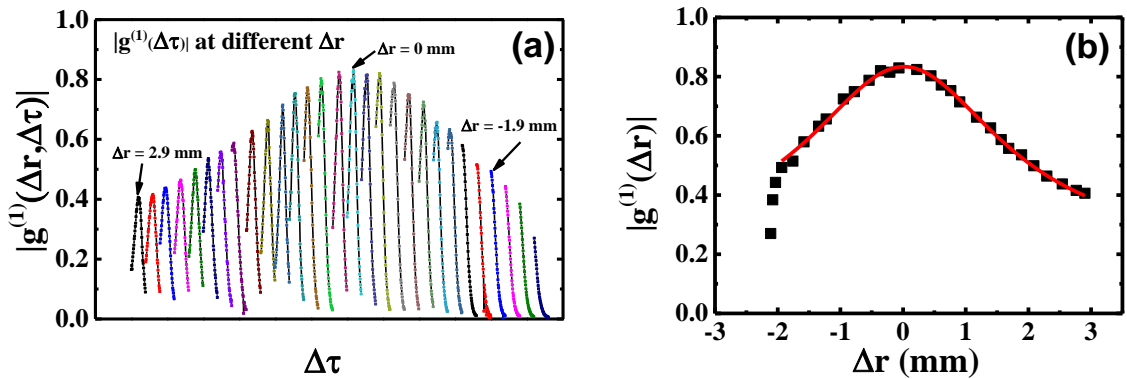

**Fig. 7.** (a) shows the calculated $|g^{(1)}(\Delta\tau)|$ plotted against fine motion of piezo ($\Delta\tau$) for different spatial separation ($\Delta r$) of $I_1$ and $I_2$. The $I_1$ beam shifts laterally from $\Delta r$ = 2.9 mm to $\Delta r$ = -2.0 mm with respect to $I_2$ beam. At each $\Delta r$, the invidual curve shows $|g^{(1)}(\Delta\tau)|$ plotted agaisnt fine $\Delta\tau$ variations of the piezo actuator. The maxima of these curves is $|g^{(1)}(\Delta r,\Delta\tau=0)|$. For each plot with different $\Delta r$, the $\Delta\tau$ is varied by ~108 fs and plots of different $\Delta r$ are manually segregated along $\Delta\tau$ axis for clarity. Therefore, scale markers are not given along $\Delta\tau$ in this plot. (b) shows $|g^{(1)}(\Delta r,\Delta\tau=0)|$ plotted agaisnt $\Delta r$ as $|g^{(1)}(\Delta r)|$. The curve is fitted with Lorentzian function and indicated by red line.

Therefore, such effect of any additional unwanted temporal phase must be corrected using temporal filtering method described above.

**IV. Conclusion**

We have modified a Michelson interferometer which can be used simultaneously for dual measurements of temporal and spatial coherence of a light source. This is specifically useful in cases where installations of double slits with variable widths and slit separations are not possible in close proximity of the source itself. Because of unwanted mixing of temporal and spatial part of coherences, it becomes difficult to estimate the nature of coherence of extended light sources. So, here we present a technique for dual measurement of coherence using a single experimental setup where we use lateral degree of freedom of one of the retroreflector to include spatial coherence measurement. We also vary the temporal delay to find the maximum of first order correlation function to estimate spatial coherence by filtering out additional temporal phase. We call this method as 'temporal filtering' in spatial coherence measurements. Similarly, another technique of 'curve overlap' was developed to extend the range of this instrument while retaining the precision. We provided detailed error analyses of these correlation measurements in section III.A.3. Therefore, this new interferometric technique can become more versatile for exploring the underlying physics of spatio-temporal coherence of an extended light source placed within a cryostat. Similar methods can be extended for measuring spatio-temporal coherences of distant light sources as well.

**Acknowledgements**

MKS is thankful to IISER-Pune for Integrated PhD Fellowship. SD acknowledges the Department of Science and Technology (DST), India (Grants # DIA/2018/000029, SR/NM/TP13/2016, CRG/2019/000412) for support. We appreciate our discussions with Dr. Sunil Nair and Dr. Shivprasad Patil of IISER-Pune.

**Data Availability**

The data that support the findings of this study are available from the corresponding author upon reasonable request(s).